\documentclass[twocolumn,floatfix,superscriptaddress,amsmath,showkeys,aps,prb,eqsecnum]{revtex4-2}
\usepackage{bm}
\usepackage{xcolor}
\usepackage{hyperref}
\usepackage{makeidx}
\usepackage{amsmath}
\usepackage{graphicx}
\usepackage{dcolumn}
\usepackage{float}
\usepackage{amsfonts}
\usepackage{amssymb}
\usepackage{color}
\usepackage{array}
\usepackage{soul}

\hypersetup{
pdfauthor = {Lopes, Vale, and Barci},
pdftitle = {Competition and coexistence of superconductivity and nematic order in a two-dimensional electron gas with quadrupolar interactions}
}

\begin{document}
\title{Competition and coexistence of superconductivity and nematic order in a two-dimensional electron gas with quadrupolar interactions}
\author{Nei Lopes}
\email{nei@cbpf.br}
\affiliation{Centro Brasileiro de Pesquisas F\'{\i}sicas, Rua Dr. Xavier Sigaud 150, Urca, 22290-180, Rio de Janeiro, Brazil}
\author{Guilherme da Silva do Vale}
\affiliation{Departamento de Física Teórica,
Universidade do Estado do Rio de Janeiro, Rua São Francisco Xavier 524, 20550-013,  
Rio de Janeiro, Brazil}
\author{Daniel G. Barci}
\affiliation{Departamento de Física Teórica,
Universidade do Estado do Rio de Janeiro, Rua São Francisco Xavier 524, 20550-013,  
Rio de Janeiro, Brazil}

\date{\today}

\begin{abstract}
We investigate the interplay between superconductivity and nematic order in a two-dimensional electron gas with competing pairing and quadrupolar forward-scattering interactions. The model includes both $s$-wave and $d$-wave superconducting channels. We compute the mean-field free energy density and determine the phase diagrams as functions of interaction strengths and temperature by solving a set of coupled self-consistent equations. At zero-temperature, we find that the nematic order competes strongly with $d$-wave superconductivity, leading to a direct first-order phase transition, while its interplay with $s$-wave pairing allows for a coexistence phase characterized by an anisotropic Fermi surface with a uniform  superconducting gap. At finite-temperatures, quadrupolar interactions promote the emergence of additional superconducting components, giving rise to regimes where $s$-wave, $d$-wave, and nematic orders coexist. Our results highlight the role of symmetry and interaction strength in shaping the phase structure and provide a minimal framework to describe intertwined nematic and superconducting phases in correlated electron systems.
\end{abstract}

\maketitle

\section{Introduction}
\label{Sec:Introduction}

Electronic quantum liquid crystal phases~\cite{Kivelson-Fradkin-Emery-1998} 
were originally introduced to describe the rich phenomenology of doped Mott insulators, 
such as the copper oxide superconductor $La_{1.6-x}Nd_{0.4}Sr_xCuO_4$. 
Subsequently, similar behavior was observed in quantum Hall systems~\cite{Lilly1999}, 
where temperature-dependent anisotropies were interpreted as the presence of  quantum liquid crystals in the ground state~~\cite{Fradkin-Kivelson1999}.   
These states of matter share fundamental symmetry properties with classical liquid crystals~\cite{deGPr1998, ChaikinLubensky2000}, 
most notably the spontaneous breaking of rotational invariance while partially preserving translational symmetry. 
For instance, quantum smectic (stripe) phases break translational symmetry along one spatial direction while remaining homogeneous along the orthogonal direction~\cite{Barci-Fradkin-2002}. 
In contrast, quantum nematic phases correspond to homogeneous (translationally invariant) Fermi fluids that spontaneously break rotational symmetry~\cite{Vadim2001}.

Following their initial identification, electronic nematic phases have been observed in a wide variety of strongly correlated materials, particularly in systems exhibiting complex phase diagrams with intertwined charge, magnetic, and superconducting orders~\cite{Fradkin2010}. 
A paradigmatic example is $Sr_3Ru_2O_7$, where pronounced magnetoresistive anisotropy near a metamagnetic quantum critical point provides strong evidence for nematic order~\cite{Grigera2001,Borzi2007}.
Subsequent studies have revealed compelling evidence for nematicity in iron-based superconductors, where nematic order appears closely connected to both magnetic and superconducting instabilities~\cite{Fernandes2014,Livanas2015,Maiti2015,Ali2018,Massat2018,Fernandes2020,Kurokawa2021,Wang2021,Mukasa2021,Rana2022,Bolmer2022,Ma2025,Yamase2013,Ishida2020,Klemm2024}. 
Similarly, in cuprate superconductors, it is now widely accepted that, under appropriate doping conditions, the normal state above the superconducting transition exhibits nematic behavior~\cite{Tranquada-2017,Taillefer-2010,Tranquada-2020}. 
More recently, the interplay between superconductivity and nematicity has also been explored in graphene-based systems~\cite{Holleis2025,Zhang2025} and Kagome materials, where superconductivity emerges in proximity to a nematic quantum critical point~\cite{Sur2023}.

From a theoretical standpoint, a central issue is understanding the competition and coexistence between superconductivity and anisotropic electronic orders. 
This problem has been extensively investigated within Hubbard-like models and related frameworks~\cite{Dixon1999,Dong2022,FengChen-2025,Zegrodnik2020}. 
In particular, nematic fluctuations have been shown to play a crucial role in stabilizing unconventional superconducting states~\cite{Barci-Fradkin-2011,Barci-Clarim-Nei-2016}.

As noted above, the term ``nematic'' has been used to identify a quantum phase that breaks rotational symmetry while preserving translational symmetry. 
However, the properties of these phases differ substantially depending on whether the original rotational group is continuous or discrete. 
For instance, the nematic phase in quantum Hall systems appears to be a fluctuating liquid crystal where the underlying lattice plays no relevant role. 
In this case, the continuous rotational symmetry is spontaneously broken, and the low-energy physics is controlled by Goldstone modes~\cite{Lawler-2006}. 
The effect of the underlying lattice is to lift the degeneracy among the otherwise equivalent vacua,  giving a small mass to the would-be Goldstone modes (pseudo-Goldstone bosons) and selecting a preferred ground state. 
The system retains approximate symmetry properties, and the explicit breaking can be treated perturbatively~\cite{Barci-Fradkin-2011}. 

In contrast, in iron-based superconductors, the tetragonal structure of the material and its deformation appear to be essential ingredients~\cite{Meese-PRL-2025,  Meese-PRB-2026} . 
Here, the continuous rotational symmetry is explicitly broken to a discrete group. 
The spontaneous breaking of this discrete group leads to the so-called ``Ising-nematic'' phase. 
In this case, there are no Goldstone modes, and the low-energy physics is typically governed by domain walls~\cite{Fernandes2020}.

From a  point of view of symmetries,   in the continuous rotational case,  the nematic order parameter is a  two-dimensional symmetric traceless tensor (two-degrees of freedom),  while in the case of Ising-Nematic,  the nematic order parameter is a single real scalar field.  

Motivated by this distinction, in this work we investigate the interplay between superconductivity and nematic order in the former scenario, where lattice effects can be treated perturbatively. 
We consider a minimal model of a two-dimensional electron gas with two competing interactions: 
(i) a quadrupolar forward-scattering interaction that, in isolation, drives an isotropic--nematic transition~\cite{Barci-Oxman-2003,Lawler-2006}, and 
(ii) a BCS-type pairing interaction~\cite{Bardeen1957, Bardeen1957a} that includes both $s$-wave and $d$-wave channels. 
The inclusion of the $d$-wave component is essential, as it shares the same rotational symmetry properties as the nematic order parameter. 
We study this simple continuous-rotation symmetric model and subsequently comment on the effect of the underlying lattice.

Within a mean-field approach, we compute the free energy density and analyze the resulting phase diagrams as functions of interaction strengths and temperature. 
The problem reduces to solving five coupled self-consistent equations corresponding to the different components of the order parameters: 
one $s$-wave superconducting component $\Delta_s$, two $d$-wave components $\Delta_d^{s,c}$, and two nematic components $\Delta_n^{s,c}$.

Our results reveal a rich interplay between superconducting and nematic orders, including regimes of competition and coexistence. 
The main findings are summarized in Figs.~\ref{fig:vdxvs}--\ref{fig:deltasxT_2}, where we present both zero-temperature and finite-temperature phase diagrams.

The paper is organized as follows: 
In Sec.~\ref{Sec:Model}, we introduce the model and derive the mean-field free energy density. 
In Sec.~\ref{Sec:results}, we present the self-consistent equations and analyze the phase diagrams for representative parameter regimes. 
Finally, in Sec.~\ref{Sec:discussions}, we summarize our results and discuss their physical implications.

\section{Model and mean-field free energy density}
\label{Sec:Model}
To investigate the interplay between superconducting and nematic orders, we consider a two-dimensional electron gas with interactions in both the particle-particle and particle-hole channels. The Hamiltonian is given by
\begin{align}
    H = \sum_{\mathbf{k} \sigma} \left(\epsilon_{\mathbf{k}}-\mu\right) c^{\dagger}_{\mathbf{k}\sigma} c_{ \mathbf{k}\sigma} 
    &+ \sum_{\mathbf{k} \mathbf{k'}}V_{\mathbf{k} \mathbf{k'}}^{SC}c^{\dagger}_{\mathbf{k}\uparrow}c^{\dagger}_{-\mathbf{k}\downarrow}c_{\mathbf{-k'}\downarrow}c_{\mathbf{k'}\uparrow} \nonumber 
    \\ &+\sum_{\mathbf{k} \mathbf{k'}}V_{\mathbf{k} \mathbf{k'}}^{N}c^{\dagger}_{\mathbf{k}\uparrow}c_{\mathbf{k}\downarrow}c^{\dagger}_{\mathbf{k'}\downarrow}c_{\mathbf{k'}\uparrow}
\label{Hamiltonian_complete}
\end{align}
where $\epsilon(\mathbf{k})$ denotes the electronic dispersion and $\mu$ is the chemical potential. For a free electron gas, the dispersion is quadratic, $\epsilon_{\mathbf{k}} = k^2/(2m)$. However, without loss of generality, more realistic band structures can be considered~\cite{Barci-Oxman-2003}. The operators $c^{\dagger}_{\mathbf{k}\sigma}$ ($c_{\mathbf{k}\sigma}$) create (annihilate) electrons with momentum $\mathbf{k}$ and spin $\sigma$. The functions $V_{\mathbf{k} \mathbf{k'}}^{SC}$ and $V_{\mathbf{k} \mathbf{k'}}^{N}$ denote the interaction form factors in the superconducting and forward-scattering channels, respectively.

The explicit form of the interactions is guided by symmetry considerations. In two dimensions, quadrupolar forward-scattering interactions are invariant under the transformation ${\bf k}\to -{\bf k}$, which characterizes nematic symmetry. This symmetry coincides with that of a $d$-wave superconducting order parameter. Accordingly, in addition to conventional $s$-wave pairing, we include a $d$-wave component in the superconducting channel. We, therefore, consider
\begin{align}
\label{interactions}
    V_{\mathbf{k}\mathbf{k}'}^{SC} &= V_s+V_df_{\mathbf{k}\mathbf{k}'} \\
    \label{VkkN}
    V_{\mathbf{k}\mathbf{k}'}^{N} &= f_2 f_{\mathbf{k}\mathbf{k}'}
\end{align}
where $V_s$ and $V_d$ are coupling constants associated with the $s$-wave and $d$-wave superconducting channels, respectively, while $f_2$ denotes the quadrupolar coupling constant, proportional to the corresponding Landau parameter in Fermi liquid theory~\cite{nozieres-1999}. 

The sign of the coupling constants determines the nature of the interactions. Attractive interactions in the superconducting channels correspond to $V_s<0$ and $V_d<0$, favoring the formation of Cooper pairs with $s$-wave and $d$-wave symmetry, respectively, while positive values describe repulsive interactions. Similarly, $f_2<0$ ($f_2>0$) corresponds to attractive (repulsive) quadrupolar interactions. Therefore, negative values of $f_2$ promote a Pomeranchuk instability~\cite{Pomeranchuk1959} toward a nematic phase, characterized by a spontaneous distortion of the Fermi surface.

The geometric factor $f_{\mathbf{k}\mathbf{k}'}$, in Eq.~\ref{interactions} and Eq.~\ref{VkkN}, depends only on the relative angle between momentum vectors~\cite{Aquino-2019} and is defined as
\begin{equation} \label{fFunctions}
\begin{aligned}
f_{\mathbf{k} \mathbf{k'}}& = \cos{\left(2\left[\theta_{\mathbf{k}} - \theta_{\mathbf{k'}}\right]\right)}  \\
&= \cos (2\theta_{\mathbf{k}}) \cos (2\theta_{\mathbf{k'}}) + \sin (2\theta_{\mathbf{k}}) \sin (2\theta_{\mathbf{k'}}) 
\\
&= f_\mathbf{k}^c f_\mathbf{k'}^c + f_\mathbf{k}^s f_\mathbf{k'}^s
\end{aligned}
\end{equation}
where we define $f_\mathbf{k}^c = \cos(2\theta_{\mathbf{k}})$ and $f_\mathbf{k}^s = \sin(2\theta_{\mathbf{k}})$, and $\theta_{\mathbf{k}}$ is the polar angle of $\mathbf{k}$ with respect to an arbitrary $k_x$ axis. The factor of $2$ ensures invariance under the nematic transformation $\theta_{\mathbf{k}} \to \theta_{\mathbf{k}} + \pi$.

Applying a mean-field (MF) decoupling to the interaction terms in Eq.~(\ref{Hamiltonian_complete}), we obtain
\begin{align}
H_{MF}& = \sum_{\mathbf{k}} (\epsilon_{\mathbf{k}} - \mu + 2\Delta_n^{s}f^s_{\mathbf{k}} + 2\Delta_n^{c}f_\mathbf{k}^c) c^{\dagger}_{\mathbf{k}} c_{ \mathbf{k}} 
\label{decoupling} \\ 
& +\sum_{\mathbf{k}} \left(\Delta_{s}+\Delta_d^{c}f_\mathbf{k}^c+\Delta_d^{s}f_\mathbf{k}^s \right) \left(c_{\mathbf{-k}\downarrow}c_{\mathbf{k}\uparrow} + c^{\dagger}_{\mathbf{k}\uparrow}c^{\dagger}_{\mathbf{-k}\downarrow}\right) 
\nonumber \\
& -\frac{(\Delta_{s})^2}{V_s}-\frac{(\Delta_d^{s})^2+(\Delta_d^{c})^2}{V_d} -\frac{(\Delta_n^{s})^2+(\Delta_n^{c})^2}{f_2}
\nonumber 
\end{align}
where 
\begin{equation}
\Delta_{s} = V_s \sum_{\mathbf{k}} \Biggl\langle c^{\dagger}_{{\mathbf{k}}\uparrow}c^{\dagger}_{{-\mathbf{k}\downarrow}} \Biggr\rangle = (\Delta_{s})^* 
\label{eq:Delta-s}
\end{equation}
is the $s$-wave superconducting order parameter, 
\begin{align}
\Delta_d^{c}&=  V_d \sum_{\mathbf{k}} \Biggl\langle f^c_{\mathbf{k}} c^{\dagger}_{\mathbf{k}\uparrow}c^{\dagger}_{-\mathbf{k}\downarrow} \Biggr\rangle = (\Delta_d^{c})^* 
\label{eq:Delta-dc}\\
\Delta_d^{s} &=  V_d \sum_{\mathbf{k}} \Biggl\langle f^s_\mathbf{k}c^{\dagger}_{\mathbf{k}\uparrow}c^{\dagger}_{\mathbf{-k}\downarrow} \Biggr\rangle = (\Delta_d^{s})^* 
\label{eq:Delta-ds}
\end{align}
are the $d$-wave components of the superconducting order parameter, and
\begin{align}
\Delta_n^{c} =  f_2 \sum_{\mathbf{k}} \Biggl \langle f^c_\mathbf{k} c^{\dagger}_{\mathbf{k}\uparrow}c_{\mathbf{k}\downarrow}  \Biggl \rangle = (\Delta_n^{c})^*
\label{eq:Delta-nc} \\
\Delta_n^{s} =  f_2 \sum_{\mathbf{k}} \Biggl \langle f^s_\mathbf{k} c^{\dagger}_{\mathbf{k}\uparrow}c_{\mathbf{k}\downarrow}  \Biggl \rangle = (\Delta_n^{s})^*
\label{eq:Delta-ns}
\end{align}
define the two components of the nematic order parameter.  For simplicity, the spin summation in the first term of Eq.~(\ref{decoupling}) has been suppressed.

It is convenient to rewrite Eq.~(\ref{decoupling}) as
\begin{align}
H_{MF} =& \sum_{\mathbf{k}} (\epsilon_{\mathbf{k}} - \mu(\theta)) c^{\dagger}_{\mathbf{k}} c_{ \mathbf{k}}  
+ \sum_{\mathbf{k}} \Delta_{\mathbf{k}}(\theta) \left(c_{-\mathbf{k}}c_{\mathbf{k}} + c^{\dagger}_{\mathbf{k}}c^{\dagger}_{-\mathbf{k}} \right)
\nonumber  \\
 &+C_1,
\label{hamiltonianMF}
\end{align}
where the effective chemical potential depends on the momentum direction and is given by
\begin{align}
\mu(\theta) =& \mu -2 \left(\Delta_n^{s}f^s_\mathbf{k} + \Delta_n^{c}f^c_{\mathbf{k}}\right) \; .
\label{eq:mu-theta} 
\end{align}
The anisotropic superconducting gap reads
\begin{align}
\Delta_{\mathbf{k}}(\theta) =& \Delta_{s} +\Delta_d^{c}f^c_{\mathbf{k}}+\Delta_d^{s}f^s_{\mathbf{k}} 
\label{eq:Delta-theta}
\end{align}
and the constant term collects the contributions of the different order parameters,
\begin{align}
C_1 =&-  \underbrace{\frac{(\Delta_{s})^2}{V_s}}_\text{$s$-wave term}-\underbrace{\frac{(\Delta_d^{s})^2 +(\Delta_d^{c})^2}{V_d}}_\text{$d$-wave terms} 
- \underbrace{\frac{(\Delta_n^{s})^2 +(\Delta_n^{c})^2}{f_2}}_\text{nematic terms}. 
\label{eq:C1}
\end{align}

Eq.~(\ref{hamiltonianMF}) highlights the close analogy with the standard mean-field BCS Hamiltonian~\cite{Bardeen1957,Bardeen1957a}. The essential difference lies in the angular dependence of both the effective chemical potential $\mu(\theta)$, Eq.~(\ref{eq:mu-theta}), and the superconducting gap $\Delta_{\mathbf{k}}(\theta)$, Eq.~(\ref{eq:Delta-theta}), reflecting the coupling between nematic and superconducting degrees of freedom.  In the limit $\Delta_n^{s,c}=0$, the conventional BCS Hamiltonian with $s$- and $d$-wave pairing is recovered.  Conversely,  for vanishing superconducting order parameters, one obtains a mean-field description of a nematic Fermi liquid~\cite{Vadim2001}.

To diagonalize $H_{MF}$ in Eq.~(\ref{hamiltonianMF}), we employ the Bogoliubov--Valatin--de Gennes transformation~\cite{Valantin1958,Bogoliubov1958, degennes-1999}, yielding 
\begin{align}
\label{Hamiltonian_diag}
H_{MF} = \sum_{\mathbf{k}} E_{\mathbf{k}}(\theta) (\alpha^{\dagger}_{\mathbf{k}}\alpha_{\mathbf{k}} + \beta^{\dagger}_{\mathbf{k}}\beta_{\mathbf{k}}) + C_1 + C_2+ C_3~,
\end{align}
where $(\alpha,\beta)^{\dagger}_{\mathbf{k}}$ and $(\alpha,\beta)_{\mathbf{k}}$ are quasiparticle operators obtained as linear combinations of the original fermionic operators. The quasiparticle dispersion is given by
\begin{equation}
E_{\mathbf{k}}(\theta) = \sqrt{\left(\epsilon_{\mathbf{k}}-\mu(\theta)\right)^2 + \Delta_{\mathbf{k}}^2(\theta)}
\label{eq:Dispersion}
\end{equation}
which incorporates the effects of both superconducting and nematic order parameters. The constants
 \begin{align}
 C_2 &=  \sum_{\bf k}\left(\epsilon_\mathbf{k} - \mu(\theta)\right),
 \label{eq:2}  \\
 C_3 &= -\sum_{\bf k} E_\mathbf{k}(\theta)
\label{eq:C3}
 \end{align} 
arise from the diagonalization procedure.

Finally, the mean-field free energy density~\cite{Santos2010, Nei2021} is given by
\begin{align}
F &= - 2T\sum_{\mathbf{k}}\ln(1 + e^{-\beta E_\mathbf{k}(\theta)}) + \sum_{\bf k}\left(\epsilon_\mathbf{k}-\mu(\theta) - E_\mathbf{k}(\theta)\right) \nonumber \\
  &-\frac{(\Delta_{s})^2}{V_s} 
-\frac{(\Delta_d^{s})^2 +(\Delta_d^{c})^2}{V_d}  - \frac{(\Delta_n^{s})^2+(\Delta_n^{c})^2}{f_2}
\label{freeEnergy}
\end{align}
where $\beta = 1/(k_B T)$, with $k_B$ the Boltzmann constant and $T$ the temperature.

\section{Phase diagrams: numerical solution of the mean-field equations}
\label{Sec:results}

We now present and discuss the numerical results obtained for both the ground-state and finite-temperature phase diagrams as functions of the relevant model parameters.

The phase diagrams are determined by numerically minimizing the free energy density, Eq.~(\ref{freeEnergy}), with respect to the order parameters. The corresponding mean-field equations follow from the stationarity conditions
\begin{equation}\label{self_consistent_eqs}
\begin{aligned}
\frac{\partial F}{\partial \Delta_{s}} = \frac{\partial F}{\partial \Delta_d^{{c}}} =
\frac{\partial F}{\partial \Delta_d^{{s}}} = \frac{\partial F}{\partial \Delta_n^{{c}}} = \frac{\partial F}{\partial \Delta_n^{{s}}} = 0 \, .
\end{aligned}
\end{equation}
After straightforward algebraic manipulations, these conditions lead to a set of five coupled self-consistent equations:
\begin{align}
\Delta_{s} &= -\frac{V_{s}}{2}\int d^2k \frac{\Delta_{\mathbf{k}}(\theta)}{E_{\mathbf{k}}(\theta)}\tanh{\left(\frac{\beta E_{\mathbf{k}}(\theta)}{2} \right)}
\label{eq:Ds} \\
\Delta_d^{c} &=-\frac{V_{d}}{2} \int d^2k  \frac{f_\mathbf{k}^c \Delta_{\mathbf{k}}(\theta)}{E_{\mathbf{k}}(\theta)}\tanh{\left(\frac{\beta E_{\mathbf{k}}(\theta)}{2} \right)}  
\label{eq:Ddc}\\
\Delta_d^{s} &=-\frac{V_{d}}{2}  \int d^2k  \frac{f_\mathbf{k}^s \Delta_{\mathbf{k}}(\theta)}{E_{\mathbf{k}}(\theta)}\tanh{\left(\frac{\beta E_{\mathbf{k}}(\theta)}{2} \right)}
\label{eq:Dds} \\
\Delta_n^{s} &=- f_2\!\! \int \!\!d^2k f_\mathbf{k}^s\left[\frac{(\epsilon_{\mathbf{k}} - \mu(\theta))\tanh{\left(\frac{\beta E_{\mathbf{k}}(\theta)}{2} \right)} }{E_{\mathbf{k}}(\theta)}- 1\right]
\label{eq:Dns} \\
\Delta_n^{c}& = -f_2 \!\! \int  \!\! d^2k f_\mathbf{k}^c \left[\frac{(\epsilon_{\mathbf{k}} - \mu(\theta))\tanh{\left(\frac{\beta E_{\mathbf{k}}(\theta)}{2} \right)}}{E_{\mathbf{k}}(\theta)} - 1 \right]
\label{eq:Dnc} 
\end{align}

The momentum integration is performed over a shell around the Fermi surface:
\begin{equation}
\int d^2k= \int_0^{2\pi} d\theta\int_{\sqrt{(1-\alpha)2m\mu(\theta)}}^{\sqrt{(1+\alpha)2m\mu(\theta)}}dk\, k \, .
\end{equation}
The parameter $\alpha$ controls the width of the shell.  Notably,  the nematic order parameter enters the integration cutoff through $\mu(\theta)$. When $\Delta_n^{c,s} \neq 0$, the Fermi surface becomes anisotropic (elliptical), and the integration domain must consistently follow this deformation.

In what follows, we explore representative regions of parameter space. Throughout, we fix $\mu = 1.0$ and $m = 1.0$, and restrict the momentum integration to a narrow shell around the Fermi surface ($\alpha=0.1$), which captures the dominant low-temperature contributions.

\subsection{Competition between $s$- and $d$-wave superconductivity at $T=0$}
\label{Sec:competingSC}

We first analyze the competition between $s$-wave and $d$-wave superconductivity at zero-temperature, neglecting quadrupolar interactions ($f_2 = 0$). In this limit, the problem reduces to solving Eqs.~(\ref{eq:Ds})--(\ref{eq:Dds}).

\begin{figure}[htb]
  \includegraphics[width=\columnwidth]{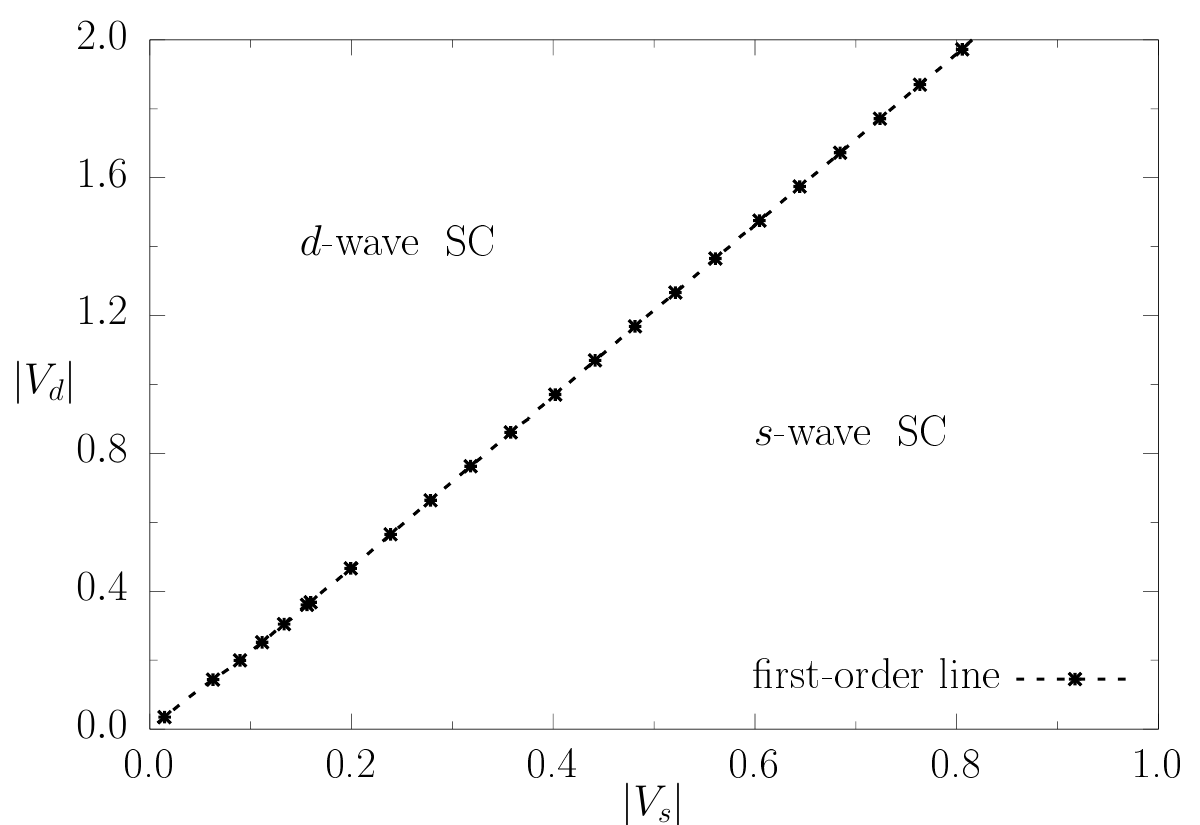}
  \caption{Phase diagram (ground state) for superconducting orders as a function of the interaction strengths $V_s$ and $V_d$ (in modulus), for $f_2 = 0.0$. The $s$- and $d$-wave superconducting phases are separated by a first-order transition line (dashed-dotted). Increasing $|V_s|$ ($|V_d|$) stabilizes the $s$-wave ($d$-wave) superconducting ground state.} 
	\label{fig:vdxvs}
\end{figure}

Fig.~\ref{fig:vdxvs} shows the ground-state phase diagram as a function of the interaction strengths $V_s$ and $V_d$. As expected, negative values with increasing $|V_s|$ ($|V_d|$) stabilize the $s$-wave ($d$-wave) superconducting phase. The two phases are mutually exclusive and separated by a first-order phase transition, as evidenced by the discontinuous behavior of the order parameters and corroborated by the corresponding free energy density analysis.

Furthermore, in the vicinity of $V_s \approx V_d$, the $s$-wave phase is energetically favored, while the $d$-wave solution persists as a metastable state. This metastability can be lifted by increasing $V_d$, even in the absence of Fermi surface anisotropy.

\subsection{$d$-wave superconductivity and nematicity at $T=0$}
\label{Sec:dwave_nem}

We now investigate the interplay between $d$-wave superconductivity and nematic order by setting $V_s = 0$ and varying $V_d$ and $f_2$. In this case, Eqs.~(\ref{eq:Ddc})--(\ref{eq:Dnc}) must be solved.
\begin{figure}[b]
  \includegraphics[width=\columnwidth]{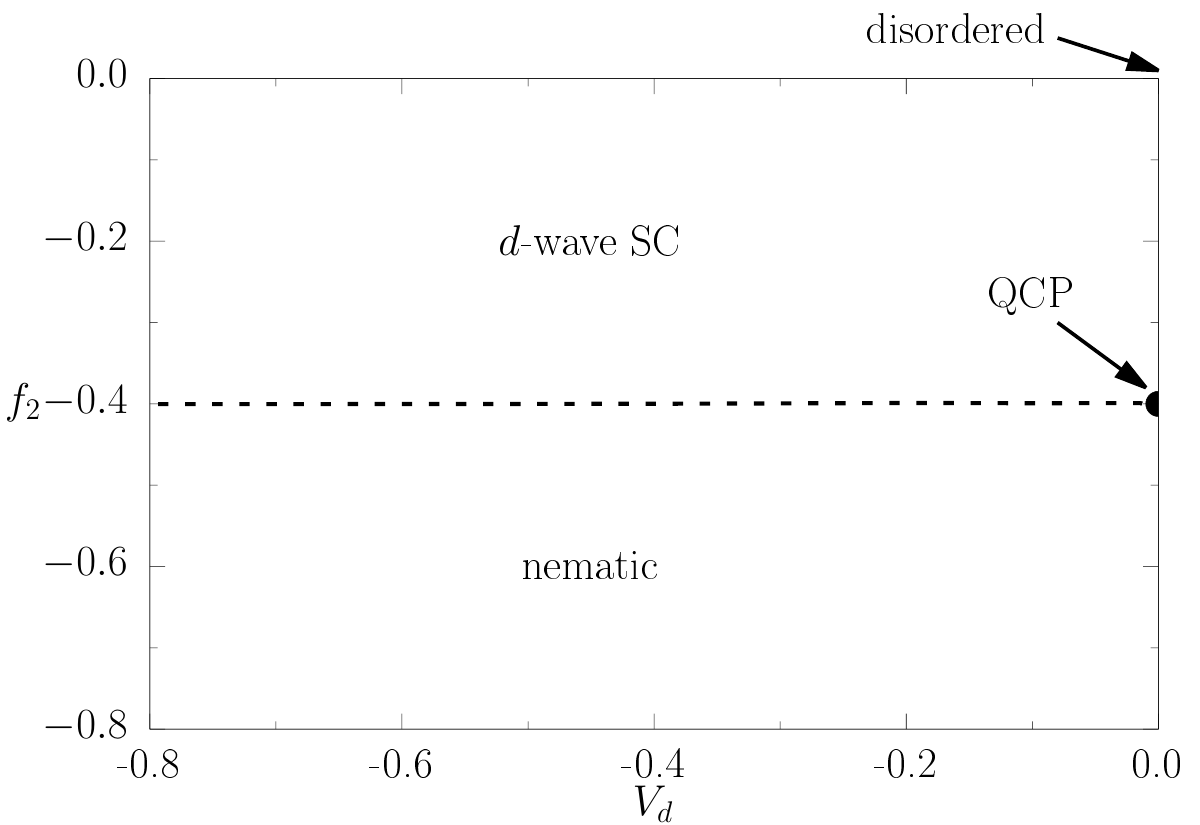}
  \caption{Phase diagram (ground state) for $d$-wave superconductivity and nematic order as a function of the interaction strengths $V_d$ and $f_2$. At $V_d = 0$, the diagram reduces to the isotropic--nematic transition. Increasing $|V_d|$ stabilizes a $d$-wave superconducting phase for $f_2 > -0.4$, while the nematic phase remains stable for $f_2 < -0.4$. The transition between the $d$-wave superconducting and nematic phases is a first-order one.} 
	\label{fig:f2xvd}
\end{figure} 

Fig.~\ref{fig:f2xvd} displays the corresponding ground state phase diagram. For $V_d=0$, the system reduces to a Fermi liquid with quadrupolar interactions~\cite{Vadim2001}. For sufficiently strong attractive quadrupolar interactions ($f_2<0$), the system undergoes an isotropic–nematic transition at a critical value $f_2^c \simeq -0.4$.  When expressed in terms of the dimensionless Landau parameter $F_2 =8N_F f_2$,  where $N_F$ is the density of state at the Fermi surface,  this value corresponds to the well-known Pomeranchuk instability\cite{Pomeranchuk1959} at $F_2 \sim -1$.   This is a quantum critical point (QCP) with dynamic critical exponent $z =3$~\cite{Lawler-2006}.  

In the presence of $d$-wave pairing ($V_d \neq 0$), the critical value $f_2^c$ remains essentially unchanged. However, it now marks a first-order phase transition between a $d$-wave superconducting phase and a nematic phase. Thus, sufficiently strong attractive  quadrupolar interactions suppress $d$-wave superconductivity discontinuously.   Therefore,  as indicated in Fig. \ref{fig:f2xvd},  the topology of the ground state phase diagram is a first-order line ending in a quantum critical point.

\subsection{$s$-wave superconductivity and nematicity at $T = 0$}
\label{Sec:swave_nem}

We next consider the interplay between $s$-wave superconductivity and nematic order by setting $V_d=0$ and solving Eqs.~(\ref{eq:Ds}), (\ref{eq:Dns}), and (\ref{eq:Dnc}).

\begin{figure}[b]
  \includegraphics[width=\columnwidth]{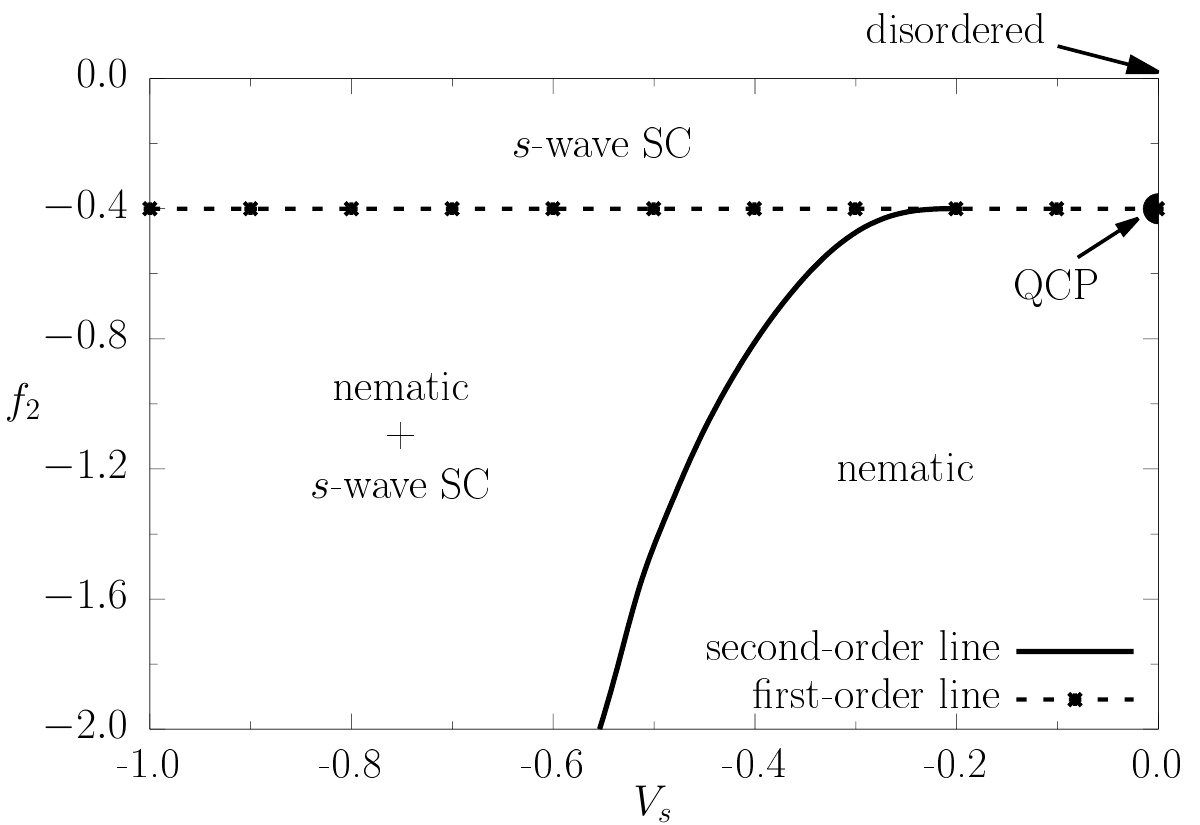}
  \caption{Phase diagram (ground state) for $s$-wave superconductivity ($\Delta_s \neq 0$) and nematic order ($\Delta_{n}^{s,c} \neq 0$) as a function of the interaction strengths $V_s$ and $f_2$. A critical quadrupolar coupling occurs at $f_2=-0.4$, separating the isotropic and nematic regimes. For small $|V_s|$, increasing $|f_2|$ stabilizes a nematic phase, while sufficiently large $|V_s|$ leads to a coexistence phase with both superconducting and nematic order. Depending on the path in parameter space, the system undergoes either first- or second-order phase transitions. Continuous (dash-dotted) lines denote second-order (first-order) transitions.} 
\label{fig:f2xvs}
\end{figure}

Fig.~\ref{fig:f2xvs} presents the zero-temperature phase diagram. For $V_s=0$, the standard isotropic–nematic transition is recovered, occurring continuously at $f_2=-0.4$.
For small $|V_s|$, an $s$-wave superconducting phase develops for $f_2>-0.4$, while sufficiently strong quadrupolar attraction suppresses superconductivity via a first-order transition to a nematic phase, in close analogy with the $d$-wave case.

However, for larger values of $|V_s|$, a qualitatively different behavior emerges. In this regime, quadrupolar interactions are unable to fully suppress superconductivity, and a coexistence phase appears in which a uniform superconducting gap develops on an anisotropic (nematic) Fermi surface. Conversely, starting from the nematic phase, increasing $|V_s|$ leads to a continuous transition into the coexistence region.

These results highlight the crucial role of symmetry: unlike the $d$-wave case, the $s$-wave and nematic order parameters transform differently under rotations, leading to a markedly distinct phase diagram topology.

\subsection{Finite-temperature effects}
\label{Sec:all}

We now consider all interaction channels and analyze finite-temperature effects by solving the complete set of coupled equations, Eqs.~(\ref{eq:Ds})--(\ref{eq:Dnc}).

\begin{figure}[b]
  \includegraphics[width=\columnwidth]{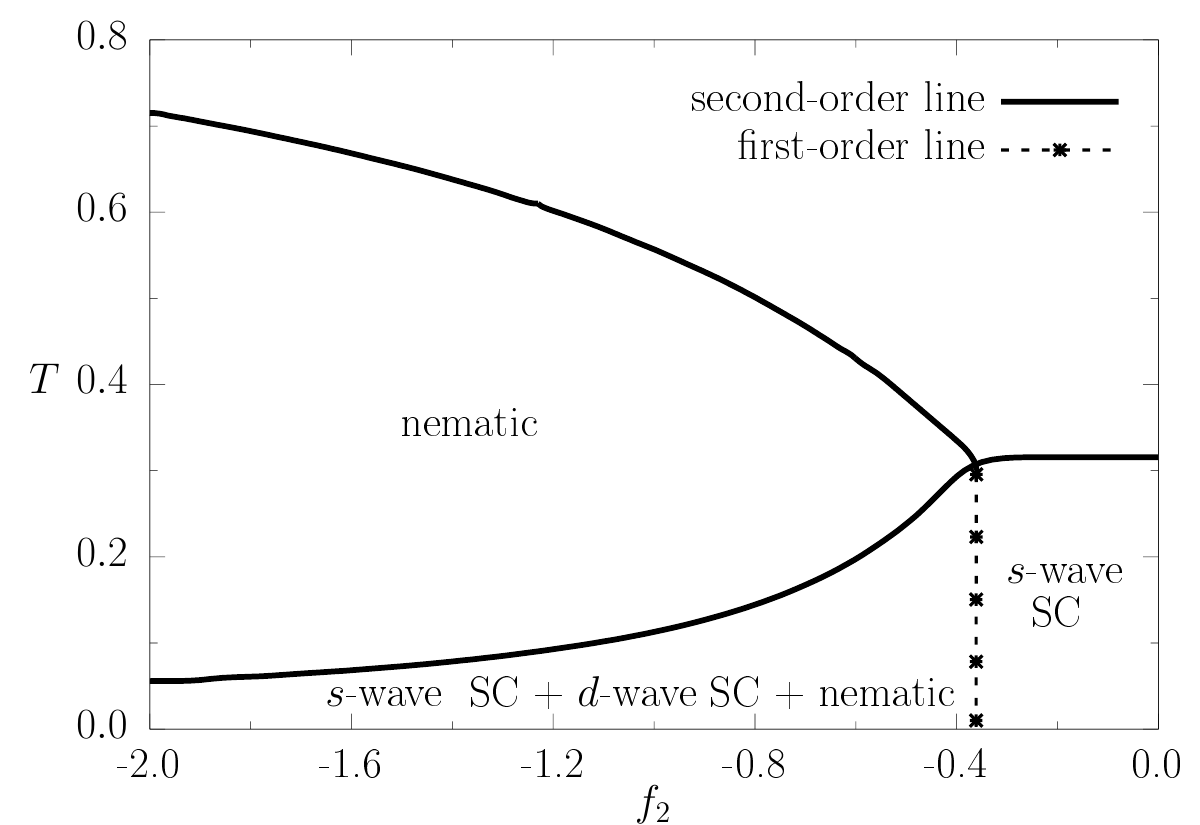}
  \caption{Critical-temperature phase diagram as a function of the quadrupolar interaction $f_2$ and temperature, including all interaction channels, for fixed $V_s = V_d = 1.0$. Continuous (dash-dotted) lines denote second-order (first-order) phase transitions. For small $|f_2|$, the ground state is an $s$-wave superconductor, consistent with Fig.~\ref{fig:vdxvs}, and the superconducting phase is suppressed with increasing temperature in the standard BCS manner. For larger $|f_2|$ and low temperatures, a first-order phase transition leads to a coexistence phase with superconducting ($s$- and $d$-wave) and nematic orders. Upon further increasing temperature, superconductivity is suppressed, yielding a purely nematic phase, which eventually undergoes a continuous transition to the disordered state at higher temperature.} 
	\label{fig:deltasxT_1}
\end{figure}

Fig.~\ref{fig:deltasxT_1} shows the phase diagram as a function of $f_2$ and temperature for $V_s = V_d = 1.0$. At $f_2=0$, the ground state is an $s$-wave superconductor, consistent with Fig.~\ref{fig:vdxvs}.
For weak quadrupolar interactions, the temperature dependence follows the conventional BCS behavior. As $|f_2|$ increases, the system undergoes a first-order phase transition into a coexistence phase where $s$-wave, $d$-wave, and nematic orders are simultaneously present. Notably, although the ground state is $s$-wave superconducting at small $f_2$, a metastable $d$-wave solution exists and becomes stabilized at larger $|f_2|$.

Upon increasing temperature, superconducting order is suppressed, leading to a purely nematic phase, which eventually undergoes a continuous transition to the disordered phase. From a topological point of view,  the phase diagram of Fig.~\ref{fig:deltasxT_1} present a tetra-critical point, where three second-order lines meet a first-order transition line. 

\begin{figure}[b]
 \includegraphics[width=\columnwidth]{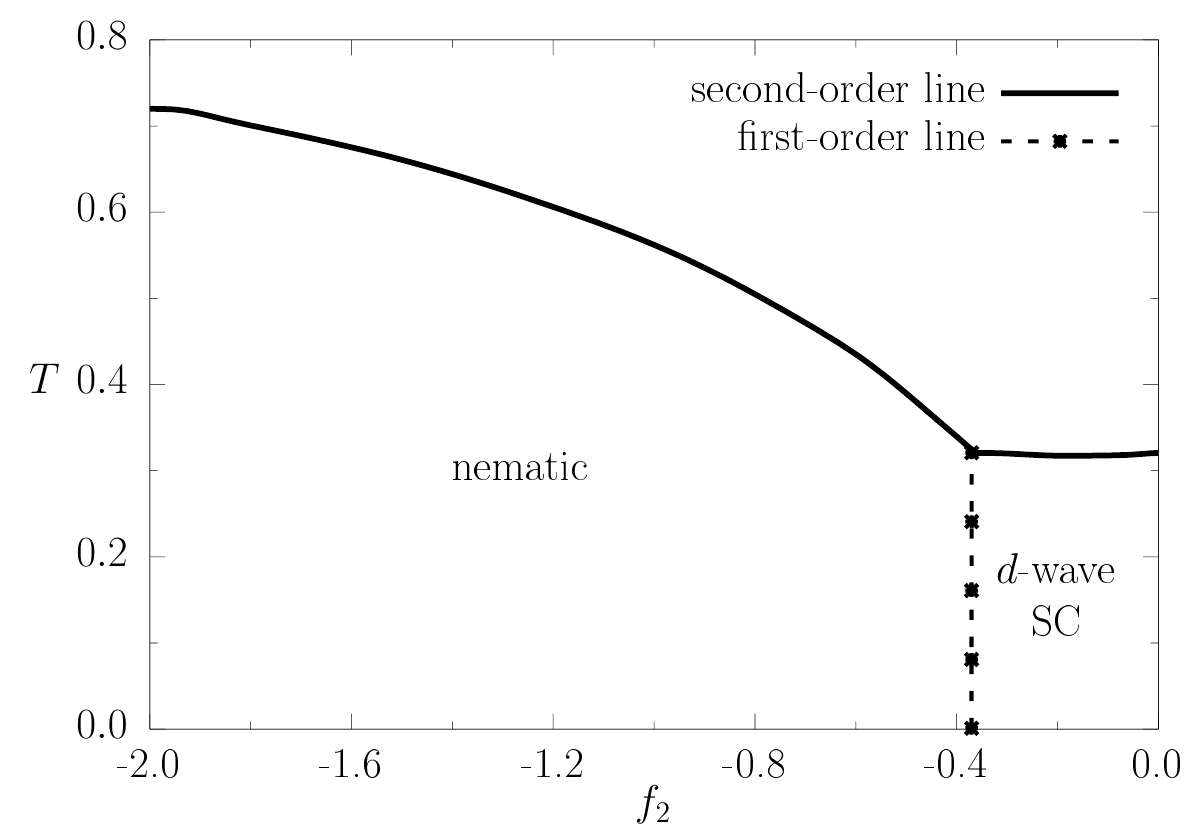}
 \caption{Critical-temperature phase diagram as a function of the quadrupolar interaction $f_2$ and temperature, including all interaction channels, for fixed $V_s = -0.2$ and $V_d = -2.0$, corresponding to a $d$-wave superconducting ground state at $f_2 = 0.0$. Continuous (dash-dotted) lines denote second-order (first-order) phase transitions. In the low-temperature regime, increasing $|f_2|$ drives a first-order transition from a $d$-wave superconducting phase to a purely nematic phase. At higher temperatures, both orders are continuously suppressed, leading to a disordered phase. }
	\label{fig:deltasxT_2}
\end{figure}

Fig.~\ref{fig:deltasxT_2} presents the complementary case where the system has a $d$-wave superconducting ground state at $f_2=0$, obtained by choosing $V_s =- 0.2$ and $V_d = -2.0$, please see Fig.~\ref{fig:vdxvs}. At low temperatures, the phase diagram is consistent with the zero-temperature results, displaying a first-order phase transition between $d$-wave superconductivity and nematic order. Increasing temperature leads to continuous transitions to the disordered phase.

In this regime, although an $s$-wave superconducting solution exists as a metastable state, it is not stabilized by quadrupolar interactions due to the mismatch in symmetry.  For this reason,  the topology of the phase diagram in Fig. \ref{fig:deltasxT_2} is different from Fig. \ref{fig:deltasxT_1}, showing a tri-critical point where two second-order phase transitions meet a first-order one.  

In all cases, phase boundaries and the nature of the transitions are determined through a detailed numerical analysis of both the order parameters and the free energy density (not shown).

In summary, our results demonstrate that quadrupolar interactions, together with the relative strength of superconducting channels, play a central role in determining the phase structure of the system. The resulting phase diagrams exhibit a rich interplay of competing and coexisting orders, particularly at low temperatures.

\section{Summary and discussions}
\label{Sec:discussions}

In this work, we have investigated the interplay between superconductivity and nematic order within a two-dimensional electron gas model that incorporates both pairing and quadrupolar forward-scattering interactions. By treating $s$-wave and $d$-wave superconducting channels on equal footing with a nematic instability, we have constructed a minimal  framework that captures the competition and coexistence of these orders within a unified mean-field description.

Our analysis of the zero-temperature phase diagrams shows that the relative symmetry of the competing order parameters plays a central role in determining the structure of the phase diagram.  In particular, when nematicity competes with $d$-wave superconductivity, both sharing the same rotational symmetry, the system exhibits a direct first-order transition between the two phases. In contrast, when nematic order competes with $s$-wave superconductivity, which belongs to a different symmetry class, a coexistence phase emerges, characterized by a uniform superconducting gap on an anisotropic (nematic) Fermi surface. These results underscore the importance of superconducting and nematic phases with different symmetries in stabilizing mixed ordered states.

At finite-temperatures, we find that quadrupolar interactions not only promote nematic order but also induce additional superconducting components. In particular, a metastable $d$-wave solution may become stabilized in the presence of sufficiently strong nematic fluctuations, leading to a regime in which $s$-wave, $d$-wave, and nematic orders coexist. Upon increasing temperature, superconducting order is progressively suppressed, giving way to a purely nematic phase, which in turn undergoes a continuous transition to the disordered state.

Overall, our results demonstrate that quadrupolar interactions provide a natural mechanism for coupling superconducting and nematic degrees of freedom, leading to rich phase diagrams that include first-order and continuous transitions, coexistence regions, and multicritical behavior. The qualitative features identified here are expected to be relevant for a broad class of strongly correlated systems in which nematicity and superconductivity are intertwined,  including cuprates and other low-dimensional materials.

It is important to emphasize that the present analysis is based on a mean-field treatment of a rotationally invariant system in two-dimensions. From a formal standpoint, long-range nematic order at finite-temperature is prohibited by the Mermin--Wagner theorem~\cite{Mermin1966}, due to the strong fluctuations associated with the corresponding Goldstone modes.  In this context, the isotropic--nematic transition in two-dimensions is expected to be of the Kosterlitz--Thouless type, driven by the unbinding of topological defects (disclinations)~\cite{Wexler-2001, Barci-Stariolo-2007,Barci-Stariolo-2009}.

However, in realistic systems, even a weak coupling to an underlying lattice explicitly breaks continuous rotational symmetry, thereby opening a gap in the spectrum of the would-be Goldstone modes~\cite{Barci-Fradkin-2011}.  Under these conditions, long-range nematic order can be stabilized, and the mean-field phase diagrams obtained here are expected to provide a qualitatively reliable description of the phase structure. This scenario, involving a fluctuating electronic liquid crystal weakly coupled to the lattice, is conceptually distinct from approaches in which rotational symmetry is explicitly broken at the outset~\cite{Fernandes2020}.

Future work should address the role of fluctuations beyond mean field, particularly near critical points,  as well as the extension of the present framework to more realistic band structures and interaction mechanisms. Such developments may provide further insight into the microscopic origin of nematicity and its interplay with unconventional superconductivity.

\section*{ACKNOWLEDGMENTS}
We acknowledge the support of the INCT project Advanced Quantum Materials, involving the Brazilian agencies CNPq (Proc. 408766/2024-7), FAPESP (Proc. 2025/27091-3), and CAPES.
The Brazilian agencies {\em Funda\c c\~ao Carlos Chagas Filho de Amparo \`a Pesquisa do Estado do Rio de Janeiro} (FAPERJ), {\em Conselho Nacional de Desenvolvimento Cient\'\i fico e Tecnol\'ogico} (CNPq) and {\em Coordena\c c\~ao  de Aperfei\c coamento de Pessoal de N\'\i vel Superior}  (CAPES) - Finance Code 001,  are acknowledged  for partial financial support.  G.S.V. would like to thank the CAPES for the M.Sc.  fellowship.


%

\end{document}